\documentclass[12pt,a4paper]{article}
\usepackage[T2A]{fontenc}
\usepackage[utf8]{inputenc}
\usepackage[english]{babel}
\usepackage{amssymb}
\usepackage{amsmath}
\usepackage{graphicx}
\usepackage[small,sc,center]{titlesec}
\usepackage{indentfirst}
\usepackage[labelsep=period]{caption}
\usepackage{caption}
\begin{document}

\begin{center}
	\textbf{Models of long--period variables of the globular cluster 47~Tuc}

	\vskip 3mm
	\textbf{Yu. A. Fadeyev\footnote{E--mail: fadeyev@inasan.ru}}

	\textit{Institute of Astronomy, Russian Academy of Sciences,
    		Pyatnitskaya ul. 48, Moscow, 119017 Russia} \\

	Received October 9, 2025; revised October 19, 2025; accepted October 19, 2025
\end{center}

\vskip 10pt
\noindent

\textbf{Abstract} ---
Stellar evolution computations were carried out for stars with a main sequence mass
$M_\mathrm{ZAMS}=0.86M_\odot$ and initial metal abundance $Z=0.003$ and $Z=0.004$.
Selected models of evolutionary sequences were used for calculation of radial pulsations
in the RGB, eAGB and TP--AGB evolutionary stages.
Not all pulsating red giants of the globular cluster 47~Tuc are shown to belong to the Mira
variables because the lower limit of pulsation periods at the TP--AGB stage is $\approx 70$~day,
whereas during the eAGB evolutionary stage the periods of radial oscillations range from
$\approx 5$ to $\approx 40$~day.
Periods and luminosities of hydrodynamic models of eAGB and TP--AGB pulsating stars locate along
the common period--luminosity relation.
Small masses of Mira variables in the globular cluster 47~Tuc ($0.54M_\odot\le M\le 0.70M_\odot$)
is the main reason for irregular large--amplitude oscillations and the dynamical instability
of outer stellar layers at pulsation periods $\Pi > 200$~day.


\section*{INTRODUCTION}

The globular cluster 47~Tuc (NGC~104) attracts attention due to its metal abundance which is nearly
by an order of magnitude higher than that of the most globular clusters.
The large opacity due to the high metal abundance seems to be the main cause of the large number
of pulsating red giants in 47~Tuc.
For example, among 117 long--period variables discovered up to 2001 in 102 globular clusters
14 variables of this type were observed in 47~Tuc, whereas the number of red pulsating variables
in other globular clusters (with exception of $\omega$~Cen) was only a few (Clement et al. 2001).
During a following quarter of a century the number of long--period variables discovered in 47~Tuc
was more than doubled (Lebzelter and Wood 2005; Lebzelter et al. 2005; Percy and Gupta 2021).
Moreover, infrared excesses detected in spectra of long--period variables of 47~Tuc indicate
condensation of dust grains in outer layers of stellar atmospheres (Origlia et al. 2002;
McDonald et al. 2011).
The main physical mechanism responsible for significant gas density increase necessary for
phase transition into solid state are the periodic shock waves accompanying the large--amplitude
stellar oscillations (Willson 2000).

At present the evolutionary status of pulsating red giants observed in 47~Tuc still remains unclear.
Lebzelter et al. (2005) noted that in the period--magnitude diagram not all pulsating variables of this
cluster obey the general relation and to explain this feature they assumed that some of these variables
ascend along the red giant branch (RGB) prior the helium flash.
To confirm this assumption Lebzelter and Wood (2005) investigated the linear pulsations of model stars
with metal abundance $Z=0.004$ and the main sequence turnoff mass $M=0.9M_\odot$.
Unfortunately, no convincing arguments of this assumption were presented because the modern estimates
of metallicity of stars in the cluster 47~Tuc are significantly smaller.

The goal of the present study is to determine the conditions of radial pulsation excitation in
red giants of the cluster 47~Tuc at the RGB stage prior to the helium flash, during the early
asymptotic giant branch (eAGB) and in the stage of thermally unstable helium shell source (TP--AGB).
The solution is based on the results of consistent calculations of stellar evolution and nonlinear
stellar pulsations where selected models of evolutionary sequences are used as initial conditions
of the Cauchy problem for equations of hydrodynamics describing stellar oscillations.
Stellar evolution was computed with the code MESA version r15140 (Paxton et al. 2019) and
details of evolutionary calculations are described in the preceding papers of the author
(Fadeyev 2023; 2024).
Equations of radiation hydrodynamics and time--dependent convection used for calculation of
stellar pulsations are discussed by Fadeyev (2013).

\section*{RESULTS OF COMPUTATIONS}

Current estimates of the age of the globular cluster 47~Tuc range from $11.8\times 10^9$ to
$12.4\times 10^9$~yr (McDonald et al. 2011; Brogaard et al. 2017; Fu et al. 2018;
Thompson et al. 2020; Simunovic et al. 2023) and in the present study we considered evolutionary
sequences of stars with the main sequence mass $M_\mathrm{ZAMS}=0.86M_\odot$ with the stellar age
at the AGB stage $\approx 11.9\times 10^9$~yr.

Computations of stellar evolution were carried out for two values of metal abundance.
The first of them is $Z=0.003$ and corresponds to the metallicity $[\textrm{Fe}/\textrm{H}] = -0.67$
provided that the solar metal abundance is $Z_\odot=0.014$ (Asplund et al. 2009).
Thus, the metal abundance $Z=0.003$ corresponds to the upper limit of metallicity range
$-0.78 \le [\textrm{Fe}/\textrm{H}] \le -0.66$ of stars in 47~Tuc
(Gratton et al. 2003; McWilliam and Bernstein, 2008; Thygesen et al. 2014).
Additional calculations for $Z=0.004$ ($[\textrm{Fe}/\textrm{H}] = -0.54$) were done
for the purpose of comparing our results with those of the work by Lebzelter and Wood (2005).

In order to evaluate the role of uncertainties in mass loss during the main sequence stage
the stellar evolution calculations were done for two parameters of the mass loss rate formula
(Reimers 1975): $\eta_\mathrm{R}=0.3$ and $\eta_\mathrm{R}=0.5$.
Therefore, the following discussion is based on hydrodynamic models of pulsating red giants that
were computed with initial conditions corresponding to the four evolutionary sequences.

\subsection*{\textsl{Models of RGB stars}}

\begin{figure}
 \centering
 \includegraphics[width=0.8\columnwidth,clip]{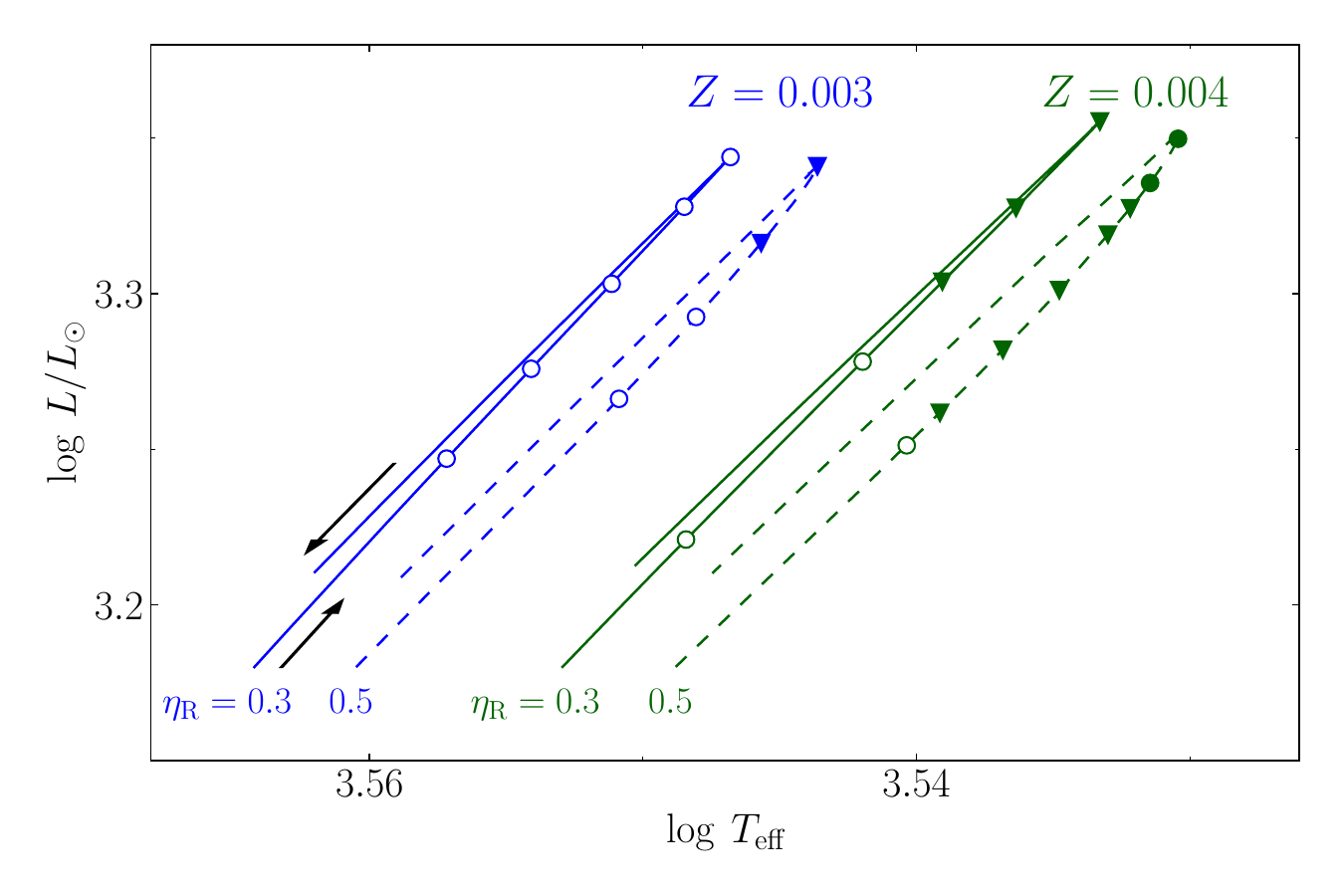}
 \caption{Evolutionary tracks near the RGB tip for the metal abundance $Z=0.003$ and $Z=0.004$
          with parameters of mass loss rate $\eta_\mathrm{R}=0.3$ (solid lines) and $\eta_\mathrm{R}=0.5$
          (dashed lines).
          Arrows indicate the direction of the star evolution along the track.
          Location of hydrodynamic models of pulsating red giants is shown by circles
          (fundamental mode pulsators) and triangles (first overtone pulsators).
          Filled symbols correspond to the models with limiting amplitude pulsations,
          whereas open symbols mark the models with decaying oscillations.}
\label{fig1}
\end{figure}

Fig.~\ref{fig1} shows the evolutionary tracks near the RGB tip.
The stellar luminosity $L$ rapidly decreases after its maximum because the hydrogen--burning shell
extinguishes and the star evolves along the descending branch hundreds times faster than
along the ascending branch.

The increase of metal abundance and mass loss rate is accompanied by displacement of evolutionary tracks
to the lower effective temperatures in the H--R diagram.
Red giants with higher $Z$ and smaller stellar mass $M$ become more unstable against radial oscillations
because of the lower gas density in the outer convection zone and more extensive ionization zones of
hydrogen and helium.
This feature is clearly seen in Fig.~\ref{fig1}, where all hydrodynamic models with
$Z=0.003$ and $\eta_\mathrm{R}=0.3$ do not pulsate, whereas increase of $Z$ and $\eta_\mathrm{R}$
is accompanied by enlargement of the pulsation instability region to lower luminosities.

Characteristics of the red giant models located near the RGB tip are listed in Table~\ref{tabl1}, where
$\Pi$, $Q$ and $\Delta R/R$ are the pulsation period, pulsation constant and relative radial displacement
amplitude at the outer boundary.
The last column of Table~\ref{tabl1} gives the duration of the pulsational instability at the RGB stage,
i.e. the time interval $\Delta t_\mathrm{ev}$ from the onset of oscillations on the ascending branch
up to the RGB tip.

\begin{table}
\caption{Characteristics of red giant models near the RGB tip}
\label{tabl1}
\begin{center}
\begin{tabular}{ccccrcll}
\hline
 $Z$ & $\eta_\mathrm{R}$ & $M/M_\odot$ & $\lg(L/L_\odot)$ & $\Pi, \textrm{day}$ & $Q, \textrm{day}$ & $\Delta R/R$ & $\Delta t_\mathrm{ev}, 10^6~\textrm{yr}$\\
\hline
 0.003 &  0.3 & 0.729 & 3.344 &  -- &   --   & 0    & 0     \\
       &  0.5 & 0.624 & 3.341 &  74 & 0.0405 & 0.10 & 0.305 \\
 0.004 &  0.3 & 0.720 & 3.355 &  76 & 0.0405 & 0.12 & 0.504 \\
       &  0.5 & 0.608 & 3.350 & 159 & 0.0773 & 0.59 & 0.761 \\
\hline
\end{tabular}
\end{center}
\end{table}

\begin{figure}
 \centering
 \includegraphics[width=0.8\columnwidth,clip]{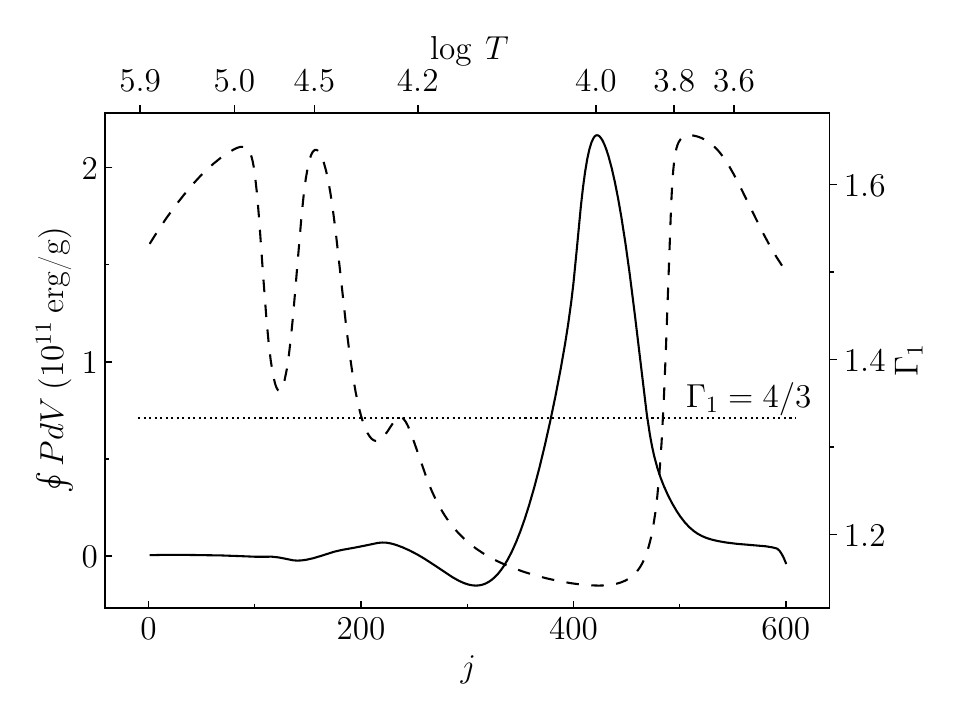}
 \caption{Specific mechanical work $\oint PdV$ done by the gas layer over the oscillation cycle
          (solid line) and the adiabatic exponent $\Gamma_1$ at the hydrostatic equilibrium (dashed line)
          versus the zone number $j$ of the hydrodynamic model.
          The temperature $T$ at the initial equilibrium conditions is given along the upper horizontal
          axis.}
\label{fig2}
\end{figure}

To get insight into contribution of stellar layers in excitation and decay of pulsational instability
we have to consider the radial dependence of the specific mechanical work $\oint PdV$ done by the
elementary spherical layer during the closed oscillation cycle, where $P$ and $V$ are the total pressure and
specific volume, respectively.
One of such relationships is shown in Fig.~\ref{fig2} for the RGB tip model of the evolutionary sequence
$Z=0.003$, $\eta_\mathrm{R}=0.5$.
For clarity Fig.~\ref{fig2} shows also the plot of the initial radial dependence of the adiabatic exponent
$\Gamma_1 = (\partial\ln P/\partial\ln V)_S$, where $S$ is the specific entropy, whereas the upper horizontal axis gives
the temperature $T$ at the initial condition of the hydrostatic and thermal equilibrium.

After attainment of the limiting amplitude the model shown in Fig.~\ref{fig2} oscillates in the first
overtone since the overtone node ($j\approx 318$) locates deeper than the inner boundary of pulsation
driving layers ($j\approx 340$).
The radius of overtone node is $r\approx 0.81\bar{R}$, where $\bar{R}$ is the mean radius of the hydrodynamic
model outer boundary.
As seen in Fig.~\ref{fig2}, main contribution into pulsation driving is provided by the zone of the partially
ionized hydrogen where the adiabatic exponent drops below the critical value $\Gamma_1 = 4/3$.

In the models of evolutionary sequences with $Z=0.004$ the zone of the partially ionized hydrogen expands
to deeper layers so that the amplitude of the limit cycle increases (see column $\Delta R/R$
in Table~\ref{tabl1}).
Moreover, near the RGB tip of the evolutionary sequence $Z=0.004$, $\eta_\mathrm{R}=0.5$
the size of the hydrogen ionization zone becomes so large that the the inner boundary of the layers
with positive mechanical work ($\oint PdV > 0$) locates deeper than the first overtone node and
the limit--cycle oscillations exist in the form of the fundamental mode.

\subsection*{\textsl{Models of eAGB stars}}

\begin{figure}
 \centering
 \includegraphics[width=0.8\columnwidth,clip]{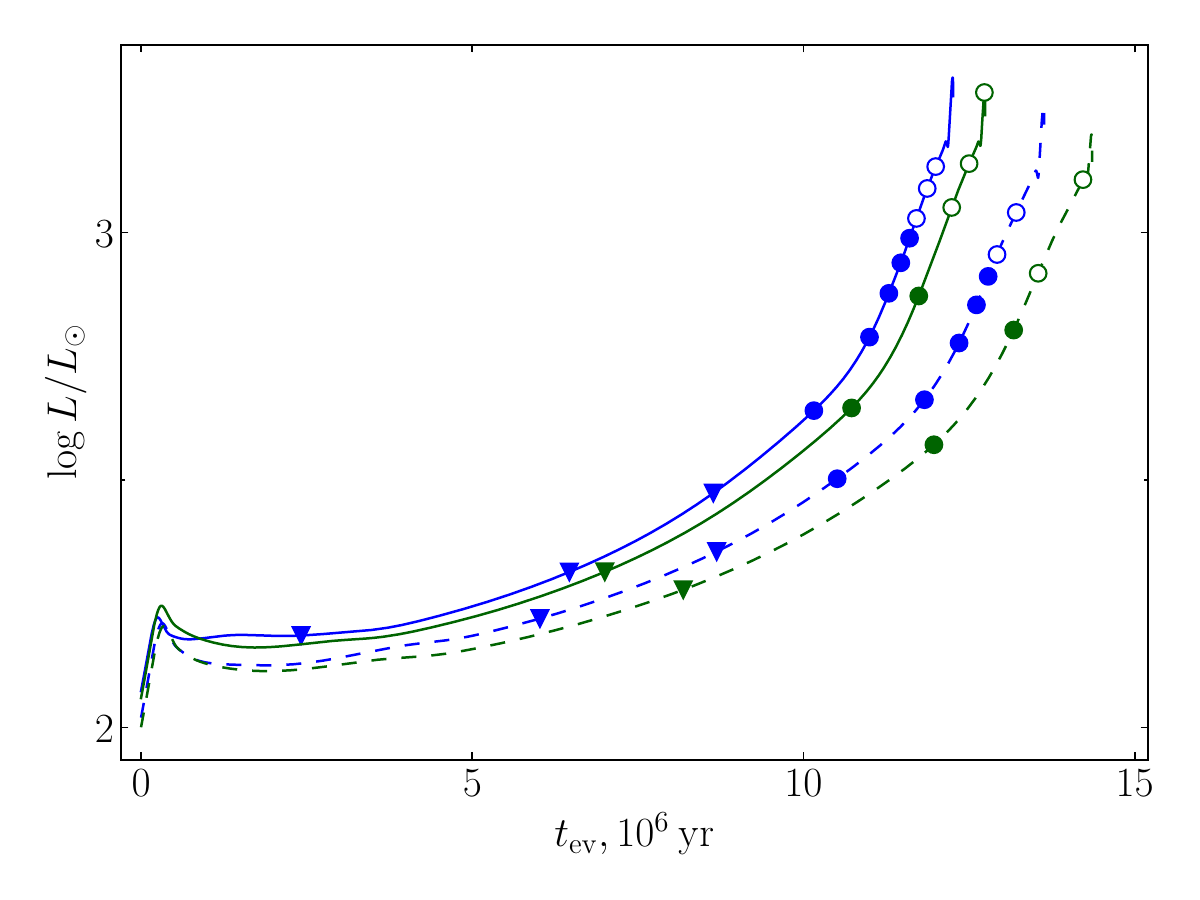}
 \caption{Evolutionary variations of the stellar luminosity during the eAGB stage.
          Designations are the same as in Fig.~\ref{fig1}.}
\label{fig3}
\end{figure}

Results of calculations for models in the eAGB stage are shown in Fig.~\ref{fig3}, where evolutionary
variations of the stellar luminosity and location of hydrodynamic models are plotted as a function of
evolution time $t_\mathrm{ev}$ with same designations as in Fig.~\ref{fig1}.
Similar to the case of RGB stars considered above the oscillations appear in the first overtone
and as the hydrogen and helium ionization zones expand due to evolutionary growth of the stellar
radius and luminosity the pulsations switch to the fundamental mode.
Pulsations of hydrodynamic models of all evolutionary sequences cease when the steadily growing
luminosity is $L\gtrsim 10^3 L_\odot$.

\begin{figure}
 \centering
 \includegraphics[width=0.8\columnwidth,clip]{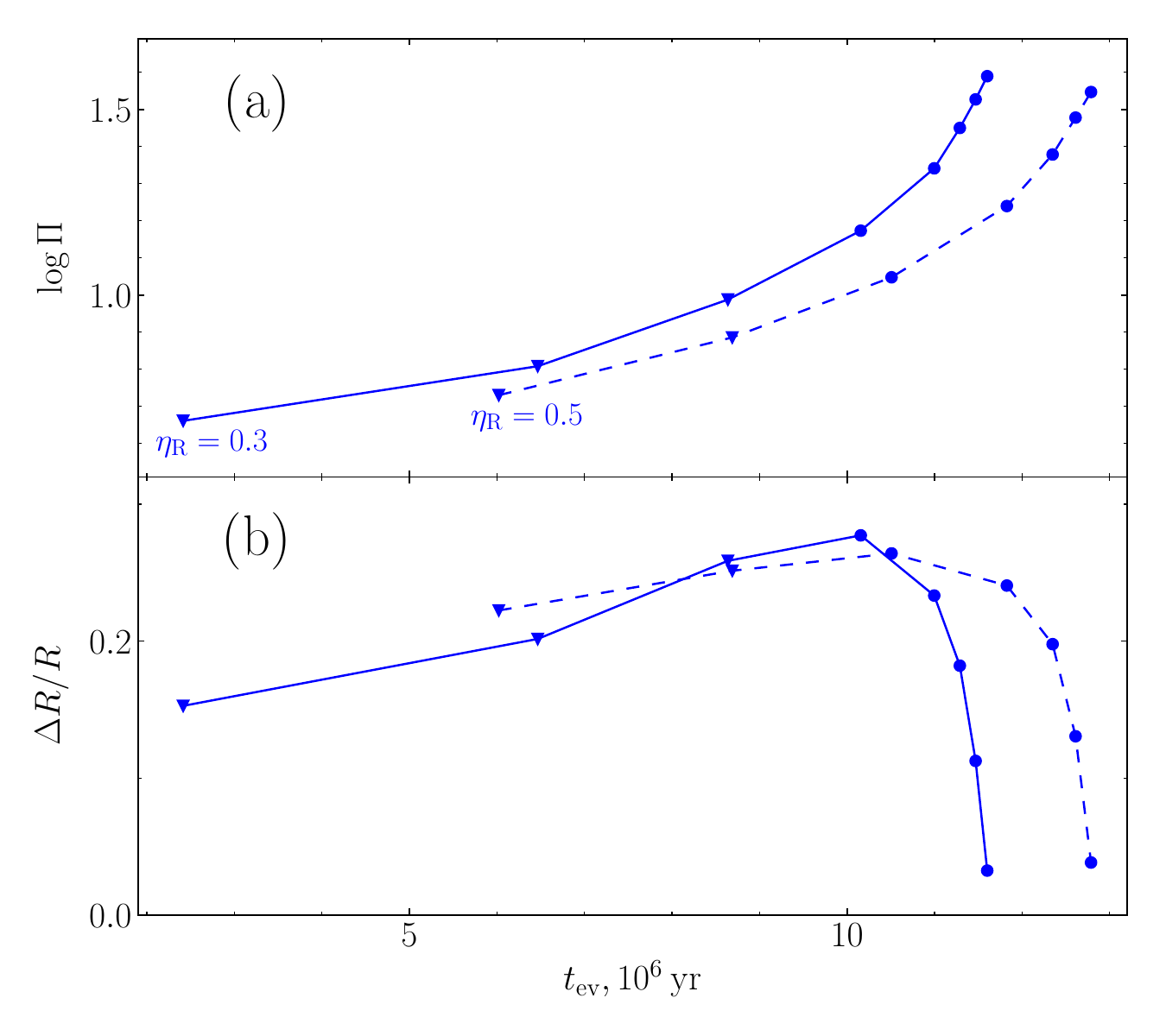}
 \caption{Evolutionary variations of the pulsation period (a) and the radial displacement amplitude
          at the outer boundary of the hydrodynamic model (b) for the metal abundance $Z=0.003$ with
          mass loss parameters $\eta_\mathrm{R}=0.3$ (solid lines) and $\eta_\mathrm{R}=0.5$ (dashed
          lines).}
\label{fig4}
\end{figure}

Fig.~\ref{fig4} shows the plots of evolutionary variations of the pulsation period $\Pi$ and
the radial displacement amplitude at the outer boundary $\Delta R/R$ for hydrodynamic models
in the eAGB stage.
The period of radial oscillations varies from $\approx 5$~day to $\approx 40$~day.
The relative amplitude of the radial displacement does not exceed $\approx 30\%$ and
stellar pulsations remain fairly regilar.

\subsection*{\textsl{Models of TP--AGB stars}}

The duration of pulsational instability of TP--AGB stars is perceptibly shorter than that of eAGB stars.
First of all, this is due to the small stellar mass.
For example, the stellar mass at the first thermal flash ranges from $0.59M_\odot$ to $0.71M_\odot$,
whereas at the onset of the post--AGB stage the mass is $M\approx 0.54M_\odot$.
The mean time interval between two successive thermal flashes is $\approx 1.6\times 10^5$~yr
so that before the onset of the post--AGB stars the star experiences only four thermal flashes.
The duration of TP--AGB stage varies from $\approx 5\times 10^5$ to $\approx 6\times 10^5$~yr.

\begin{figure}
 \centering
 \includegraphics[width=0.8\columnwidth,clip]{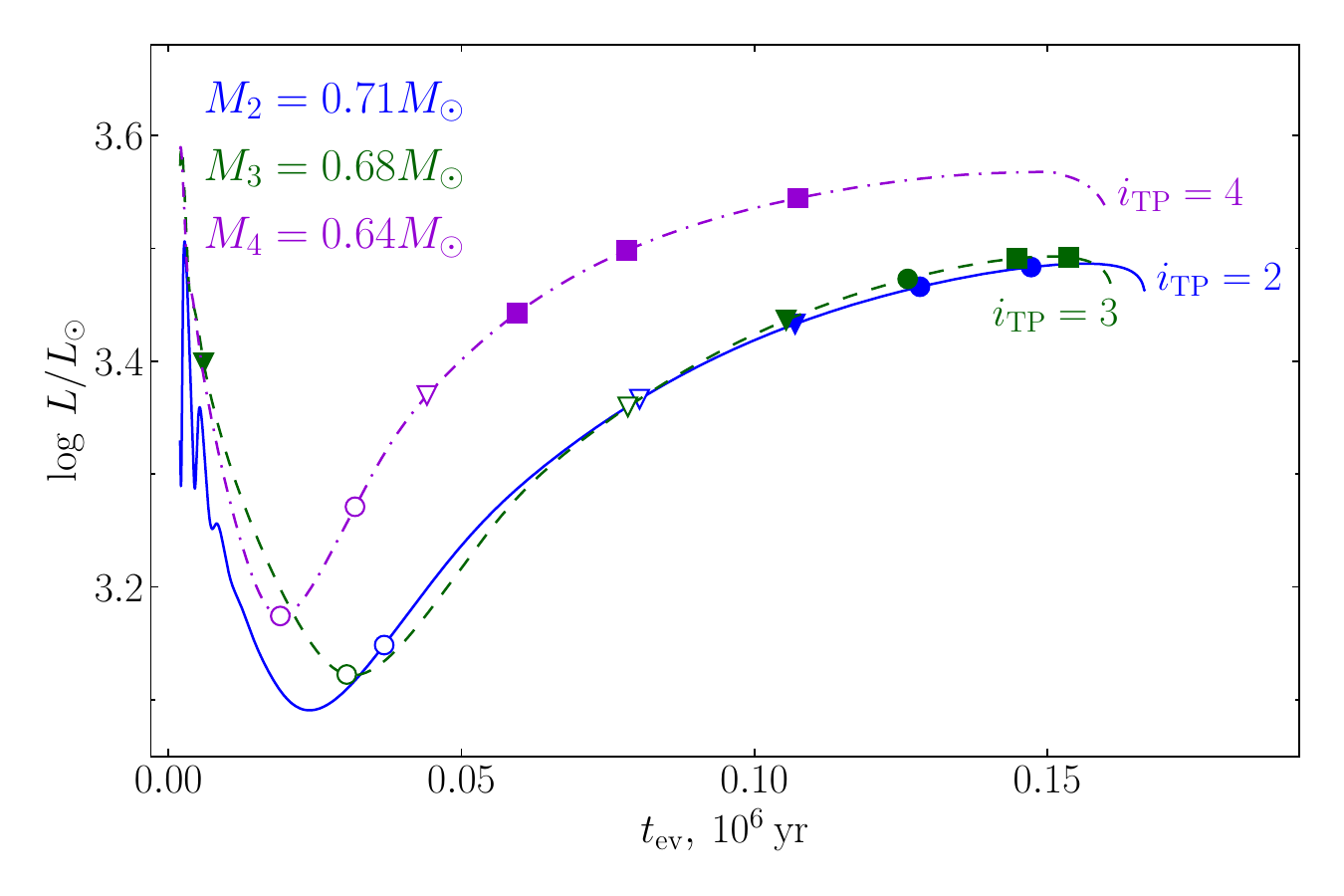}
 \caption{Luminosity variations in stars of the evolutionary sequence $Z=0.003$, $\eta_\mathrm{R}=0.3$
          after thermal flashes $2\le i_\mathrm{TP}\le 4$.
          Hydrodynamic models pulsating in the first overtone and fundamental mode are marked
          by triangles and circles, respectively.
          Models with irregular large amplitude oscillations are marked by squares.
          Open symbols correspond to models with decaying oscillations.
          The stellar masses correspond to $t_\mathrm{ev}=0$.}
\label{fig5}
\end{figure}

During the first interflash interval the red giants do not pulsate because all hydrodynamic
models show decaying oscillations.
Fig.~\ref{fig5} shows the plots of luminosity variations after thermal flashes $i_\mathrm{TP}=2$, 3 and 4
in the models of the evolutionary sequence $Z=0.003$, $\eta_\mathrm{R}=0.3$.
As seen in the plots, the radial pulsations appear when the stellar luminosity is
$\lg( L/L_\odot) \ge 3.4$ so that during significant fraction of the time interval between two
successive flashes the star does not oscillate.

Typical periods of radial oscillations during the TP--AGB stage range from $\approx 70$ to $\approx 240$~day.
Transition to irregular oscillations when the amplitude of the surface radial displacement is
$\Delta R/R \approx 1$ takes place at $\Pi\approx 170$~day.
In the plot for $i_\mathrm{TP}=4$ we marked only two hydrodynamic models with irregular
large amplitude oscillations.
The mass and luminosity of the last hydrodynamic model are $M=0.62M_\odot$ and $L=3150L_\odot$ whereas the
remaining life time in the TP--AGB stage does not exceed $\approx 1.5\times 10^4$~yr.

\section*{PERIOD--LUMINOSITY RELATION}

\begin{figure}
 \centering
 \includegraphics[width=0.8\columnwidth,clip]{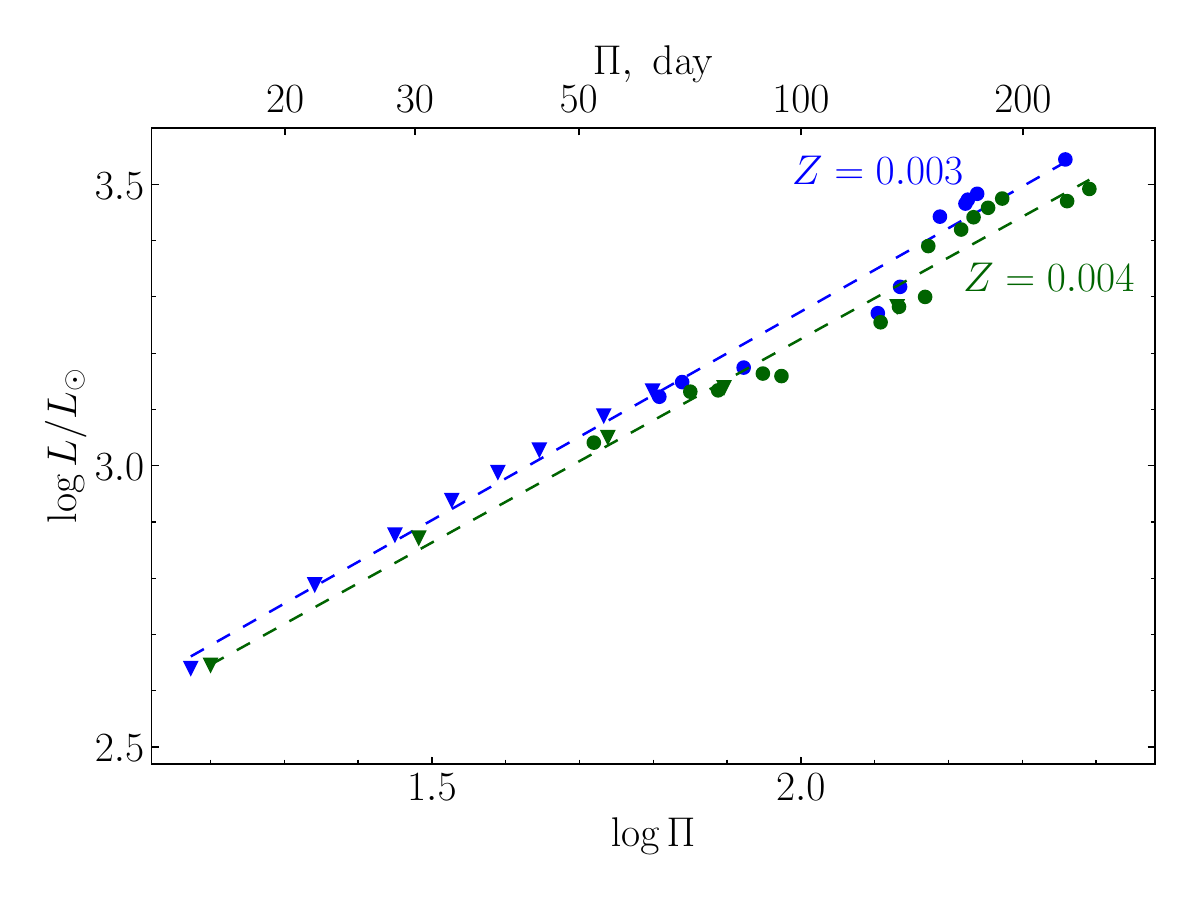}
 \caption{Period--luminosity relations for evolutionary sequences $Z=0.003$ and $Z=0.004$
          with the parameter $\eta_\mathrm{R}=0.03$.
          Hydrodynamic models of eAGB and TP--AGB stars are shown by triangles and cicrles, respectively.
          Dashed lines represent relations (\ref{pl1}) и (\ref{pl2}) for $Z=0.003$ and $Z=0.004$,
          respectfully.}
\label{fig6}
\end{figure}

Fig.~\ref{fig6} shows hydrodynamic models of evolutionary sequences $Z=0.003$ и $Z=0.004$
computed with $\eta_\mathrm{R}=0.3$ in the period--luminosity diagram.
As seen from the plots, for each $Z$ the hydrodynamic models of eAGB and TP--AGB stars locate
along the common regression line.
The corresponding period--luminosity relations are written as
\begin{gather}
\label{pl1}
\log(L/L_\odot) = 0.743\log\Pi + 1.79 \quad (Z=0.003),
\\
\label{pl2}
\log(L/L_\odot) = 0.724\log\Pi + 1.78 \quad (Z=0.004) ,
\end{gather}
where the standard error of the coefficient in the first term of right--hand sides of
(\ref{pl1}) and (\ref{pl2}) does not exceed 3\%.
It should be noted that perceptible scatter along the regression lines for $\log\Pi\gtrsim 2$
is due to the evolutionary growth of the carbon core mass and stellar luminosity which is
responsible for the scatter of points in the period--luminosity relation (Fadeyev 2024).

\section*{CONCLUSION}

Results of the consistent stellar evolution and nonlinear stellar pulsation calculations allow us to
conclude that the most of pulsating red giants in the globular cluster 47~Tuc are in the eAGB
evolutionary stage.
Excitation of radial oscillations is the same as in Mira variables but the duration
of the eAGB evolutionary stage ($\approx 1.2\times 10^7$~yr)  nearly two times longer than that
of TP--AGB stage ($\approx 6\times 10^6$~лет).
This is due to the small number of thermal flashes ($i_\mathrm{TP}\le 4$) during the TP--AGB stage
as well as due to the pulsational instability of red giants only in the end of each
interflash interval.
Dynamical instability of outer layers of hydrodynamic models with periods $\Pi\gtrsim 200$~day
implies that the principal mechanism responsible for condensation of dust grains
in red giants of 47~Tuc are periodic shock waves accompanying large--amplitude oscillations.

The lower limit of pulsation periods of hydrodynamic models in the TP--AGB stage ($\Pi\approx 70$~dat)
does not differ significantly from that of Galactic Mira variables (Samus' et al. 2017).
Moreover, the existence of red giants in 47~Tuc with periods shorter than 70 day cannot be explained
by stellar pulsations of stars near the RGB tip since at the metal abundance $Z=0.003$ the region of
the pulsational instability does not exist or remains too narrow with periods $\gtrsim 70$~day.
Asumption on higher metal abundance ($Z=0.004$) is beyond the range of observational estimates
of metallicity stars of 47~Tuc and at the same time does not allow us to resolve the problem
because the lower period limit is as small as $\Pi\approx 60$~day.
Therefore, red variables of 47~Tuc pulsating with periods shorter than 70~day are the stars
in the eAGB stage which precedes the stage of thermal instability in the helium shell source.

\section*{REFERENCES}

\begin{enumerate}
\item M.~Asplund, N.~Grevesse, A.J.~Sauval, and P.~Scott, Ann. Rev. Astron. Astrophys. \textbf{47}, 481 (2009).

\item K.~Brogaard, D.A.~VandenBerg, L.R.~Bedin, A.P.~Milone, A.~Thygesen, and F.~Grundahl, MNRAS \textbf{468}, 645 (2017).

\item C.M.~Clement, A.~Muzzin, Q.~Dufton, T.~Ponnampalam, J.~Wang, J.~Burford, A.~Richardson, T.~Rosebery, J.~Rowe,
      and H.S.~Hogg, Astron.J. \textbf{122}, 2587 (2001).

\item Yu.A. Fadeyev, Astron. Lett. \textbf{39}, 306 (2013).

\item Yu.A.~Fadeyev, Astron. Lett. \textbf{49}, 722 (2023).

\item Yu.A.~Fadeyev, Astron. Lett. \textbf{50}, 561 (2024).

\item X.~Fu, A.~Bressan, P.~Marigo, L.~Girardi, J.~Montalb\'an, Y.~Chen, and A.~Nanni, MNRAS \textbf{476}, 496 (2018).

\item R.G.~Gratton, A.~Bragaglia, E.~Carretta, G.~Clementini, S.~Desidera, F.~Grundahl, and S.~Lucatello,
      Astron. Astrophys. \textbf{408}, 529 (2003).

\item T.~Lebzelter and P.R.~Wood, Astron. Astrophys. \textbf{441}, 1117 (2005).

\item T.~Lebzelter, P.R.~Wood, K.H.~Hinkle, R.R.~Joyce, and F.C.Fekel, Astron. Astrophys. \textbf{432}, 207 (2005).

\item I.~McDonald, M.L.~Boyer, J.Th.~van~Loon, A.A.~Zijlstra, J.L.~Hora, B.~Babler, M.~Block, K.~Gordon, M.~Meade,
      M.~Meixner, K.~Misselt, T.~Robitaille, M.~Sewi{\l}o, B.~Shiao, and B.~Whitney,
      Astrophys. J. Suppl. Ser. \textrm{193}, 23 (2011).

\item A.~McWilliam and R.A.~Bernstein, Astrophys. J. \textbf{684}, 326 (2008).

\item L.~Origlia, F.R.~Ferraro, F.~Fusi~Pecci, and R.T.~Rood, Astrophys. J. \textbf{571}, 458 (2002).

\item B. Paxton, R. Smolec, J. Schwab, A. Gautschy, L. Bildsten, M. Cantiello, A. Dotter, R. Farmer, J.A. Goldberg,
      A.S. Jermyn, S.M. Kanbur, P. Marchant, A. Thoul, R.H.D. Townsend, W.M. Wolf, M. Zhang, and F.X. Timmes,
      Astrophys. J. Suppl. Ser. \textbf{243}, 10 (2019).

\item J.R.~Percy and P.~Gupta, J. Am. Associat. Var. Star Observ. \textbf{49}, 209 (2021).

\item D.~Reimers, \textit{Problems in Stellar Atmospheres and Envelopes},
       Ed. by Ed. B. Baschek, W. H. Kegel, and G. Traving (New York: Springer-Verlag, 1975), p. 229.

\item N.N.~Samus, E.V.~Kazarovets, O.V.~Durlevich, N.N.~Kireeva, and E.N.~Pastukhova, Astron. Rep. \textbf{61}, 80 (2017).

\item M.~Simunovic, T.H.~Puzia, B.~Mille, E.R.~Carrasco, A.~Dotter, S.~Cassisi, S.~Monty, and P.~Stetson,
      Astrophys. J. \textbf{950}, 135 (2023).

\item A.J.~Thygesen, L.~Sbordone, S.~Andrievsky, S.~Korotin, D.~Yong, S.~Zaggia,
       H.--G.~Ludwig, R.~Collet, M.~Asplund, P.~Ventura, F.~D'Antona, J.~Mel\'endez, and A.~D'Ercole,
       Astron. Astrophys. \textbf{572}, A108 (2014).

\item I.B.~Thompson, A.~Udalski, A.~Dotter, M.~Rozyczka, A.~Schwarzenberg--Czerny, W.~Pych, Y.~Beletsky, G.S.~Burley,
      J.L.~Marshall, A.~McWilliam, N.~Morrell, D.~Osip, A.~Monson, S.E.~Persson, M.K.~Szyma\'nski, I.~Soszy\'nski,
      R.~Poleski, K.~Ulaczyk, {\L}~Wyrzykowski, S.~Koz{\l}owski, P.~Mroz, and P.~Pietrukowicz,
      MNRAS \textbf{492}, 4254 (2020).

\item L.A.~Willson,  Ann. Rev. Astron. Astrophys. \textbf{38}, 573 (2000).

\end{enumerate}

\end{document}